\documentclass[11pt]{article}
\usepackage{a4}

\usepackage{amsfonts}
\usepackage{amsmath}
\usepackage{amssymb}

\newcommand{\bea}{\begin{eqnarray}}
\newcommand{\eea}{\end{eqnarray}}
\newcommand{\be}{\begin{equation}}
\newcommand{\ee}{\end{equation}}
\newcommand{\pa}{\partial}
\newcommand{\nn}{\nonumber}
\newcommand{\tQ}{{Q}^{NL}}
\newcommand{\tq}{{q}^{NL}}
\newcommand{\QL}{Q^L}
\newcommand{\ql}{q^L}
\newcommand{\la}{\langle}
\newcommand{\ra}{\rangle}
\renewcommand{\a}{\alpha}
\renewcommand{\b}{\beta}
\newcommand{\e}{\epsilon}
\newcommand{\ve}{\varepsilon}
\newcommand{\s}{\sigma}
\newcommand{\C}{\odot}

\makeatletter

\@addtoreset{equation}{section}
\makeatother

\def\href#1#2{#2}

\begin{document}

\begin{titlepage}

\begin{center}

\hfill 
\vskip 1.1in

\textbf{\Large M-Theory %Space-Time 
Superalgebra 
From Multiple Membranes}

\vskip .9in
{\large
Kazuyuki Furuuchi${}^{a,}$\footnote{e-mail address: furuuchi@phys.cts.nthu.edu.tw},
%Pei-Ming Ho\footnote{e-mail address: pmho@phys.ntu.edu.tw}, 
Sheng-Yu Darren Shih$^b$, %\\ \vskip 1mm  %\footnote{e-mail address: },\\ \vskip 1mm  
Tomohisa Takimi$^{c,}$\footnote{e-mail address: tomotakimi@mail.nctu.edu.tw}\\
\vskip 8mm
{\it\large
${}^a$%
National Center for Theoretical Sciences, \\
National Tsing-Hua University, Hsinchu 30013, Taiwan,
R.O.C.}\\
\vskip 2mm
{\it\large
${}^b$%
Department of Physics and Center for Theoretical Sciences, \\
National Taiwan University, Taipei 10617, Taiwan,
R.O.C.}\\
\vskip 2mm
{\it\large
${}^c$%
Department of Electrophysics, \\
National Chiao-Tung University,
Hsinchu 30010, Taiwan,
R.O.C.}\\
\noindent{ \smallskip }\\
}
\vspace{10pt}
%\maketitle
\end{center}
\begin{abstract}
We investigate space-time
supersymmetry of the model of
multiple M2-branes
proposed by
Bagger-Lambert and Gustavsson.
When there is a central element in Lie 3-algebra,
the model possesses an extra symmetry shifting 
the fermions in the central element.
Together with the original worldvolume
supersymmetry transformation,
we construct major part of 
the eleven dimensional space-time super-Poincar\'{e} 
algebra with central extensions.
Implications to transverse five-branes 
in the matrix model for M-theory are also discussed.
\end{abstract}

%\pacs{}%]
\end{titlepage}

\setcounter{footnote}{0}

\section{Introduction}

Bagger, Lambert \cite{Bagger:2006sk,Bagger:2007jr,Bagger:2007vi} 
and Gustavsson \cite{Gustavsson:2007vu,Gustavsson:2008dy}
discovered an interesting model 
for multiple M2-branes 
(which we will call BLG model in the following)
based on an algebraic structure
called Lie 3-algebra.
Since membranes are expected to be the
fundamental building blocks of 
M-theory, 
it is intriguing to ask
how much does the BLG model know about M-theory.
Important information of M-theory is contained in the structure
of the eleven dimensional space-time superalgebra, or
``M-theory superalgebra" \cite{Townsend:1995gp}.
The BLG model is not
%maifestly 
space-time supersymmetric, at least manifestly.
However, since fundamental membrane action
is expected to have space-time supersymmetry,
we may hope that
the BLG model can be related to a gauge-fixed form of
some manifestly space-time supersymmetric
formulation.

In this paper we show that the most part of
the eleven dimensional
space-time super-Poincar\'e algebra 
with central extensions
can actually be constructed
from the BLG model, and 
indeed it captures important aspects of M-theory;
namely charges of BPS branes.
One of the crucial ingredients in
constructing the space-time superalgebra is
an existence of a central element in the Lie 3-algebra
which the BLG model is based on.
The shift of bosonic as well as fermionic fields
in this central element are symmetries of the BLG model.
The shift of the bosonic fields corresponds to 
translations in space-time, whereas
the shift in the fermionic fields
represents non-linearly realized part of the
%provides a necessary piece of
space-time super-Poincar\'e algebra.

Similar discussions on the worldvolume supersymmetry algebra 
of the BLG model
which is identified with the linearly realized part of
the space-time supersymmetry
%along similar line of ours has been discussed 
can be found in 
a recent paper \cite{Passerini:2008qt}.
%\footnote{%
%We originally planned to investigate this part also,
%but the paper \cite{Passerini:2008qt}
%appeared while our results were still preliminary.}
%For this part, 
We extend the results 
by including
configurations 
which take values in
non-trace elements
(trace elements and non-trace elements are
defined in section \ref{BLGmodel})
and obtained more central charges
which provide necessary pieces
of the M-theory superalgebra.
%showed 
%that those central charges
%which were dropped in
%\cite{Passerini:2008qt}
%because they vanish for
%trace elements
%actually survive and give important pieces
%of the M-theory superalgebra.
The algebra and the central charges 
which arise by including 
the 
fermionic shift symmetry 
%were not
%considered in \cite{Passerini:2008qt} 
%and these 
are our new results.
One of our main interests is on %constructing
the charge of the five-brane
constructed in \cite{Ho:2008nn,Ho:2008ve}, 
and they are obtained only
by including the fermionic shift symmetry in the algebra.
Five-brane charges are of particular interests %interesting
because in the matrix model for M-theory 
\cite{Banks:1996vh}
transverse five-branes
are not seen 
in the superalgebra
\cite{Banks:1996nn}.

Space-time superalgebra of 
a deformed BLG model without central extensions
was constructed in \cite{Gomis:2008cv}.
Other aspects of BPS configurations
for the worldvolume supersymmetry of the BLG model 
were studied in \cite{Hosomichi:2008qk,Jeon:2008bx,Krishnan:2008zm}.

\section{Space-time superalgebra from multiple membranes}

\subsection{The Bagger-Lambert-Gustavsson model}
\label{BLGmodel}

The Bagger-Lambert action %LG model 
which was 
proposed as a description for multiple M2-branes 
\cite{Bagger:2007jr} 
(see also \cite{Bagger:2006sk,Bagger:2007vi,Gustavsson:2007vu,Gustavsson:2008dy})
has ${\cal N} = 8$ worldvolume supersymmetry.
Furthermore, it has a novel gauge symmetry based on an algebraic structure 
called Lie 3-algebra \cite{Filippov}.
For a linear space 
${\cal V} = \sum_{a=1}^{\dim {\cal V}} v_a T^a; v_a \in \mathbb{C}$,
Lie 3-algebra structure is defined by a multi-linear map
which we call 3-bracket
$[*,*,*]$ : ${\cal V}^{\otimes 3} \rightarrow {\cal V}$
satisfying following properties:\\
\ \\ 
1. Skew-symmetry:
\be
 \label{skew}
[A_{\s(1)}, A_{\s(2)} , A_{\s(3)}] = (-1)^{|\s|} [A_1, A_2, A_3].
\ee
2. Fundamental identity:
\bea 
 \label{FI} 
&&[A_1, A_2, [B_1, B_2, B_3]] \nn \\
&=& [[A_1,A_2,B_1],B_2,B_3] + [B_1,[A_1,A_2,B_2],B_3] + [B_1,B_2,[A_1,A_2,B_3]].\nn\\
\eea
A linear space endowed with a Lie 3-algebra structure
will be called Lie 3-algebra and typically denoted as ${\cal A}$
in this paper.
In terms of the basis $T^a$, Lie 3-algebra can be expressed in terms of
the structure constants $f^{abc}{}_d$:
\bea
 \label{st}
[T^a,T^b,T^c] = f^{abc}{}_d T^d .
\eea
An element 
$T^a \in {\cal A}$ is called a center if $[T^a,T^b,T^c]=0,
\forall\, T^b,T^c\in{\cal A}$,
and $f^{abc}{}_d = 0$ in this case.
To construct the action, we will also need an inner product
in Lie 3-algebra.
We assume the structure
${\cal V} = {\cal V}_{tr} \oplus {\cal V}_{ntr}$,
where elements in ${\cal V}_{tr}$
have inner product and 
elements in ${\cal V}_{ntr}$ do not.
We will refer to the elements in ${\cal V}_{tr}$
as trace elements, 
%and set of all trace elements
%as trace class.
%Similary, 
and elements in ${\cal V}_{ntr}$ 
as non-trace elements. %,
%and set of all non-trace elements as non-trace class.
By definition, 
elements $T^a, T^b \in {\cal V}_{tr}$ have
inner product $\la *, *\ra$:
${\cal V}_{tr} \otimes {\cal V}_{tr} \rightarrow {\mathbb{C}}$:
\bea
\la T^a,T^b \ra = h^{ab} .
\eea
We will call $h^{ab}$ as metric of Lie 3-algebra. %,
%and sometimes call $T^a$ as generators of Lie 3-algebra.
%
We require following invariance of the inner product
which
is important for the gauge invariance of the Bagger-Lambert action:
\bea
 \label{invm}
\la [T^a, T^b, T^c], T^d \ra 
+ \la T^c, [T^a, T^b, T^d] \ra = 0.
\eea
%We may regard $[\a,\b,*]$ as an analogue of derivation
%
Together with the skew-symmetry property (\ref{skew}),
the invariance of the metric
(\ref{invm}) requires the indices of structure constants
$f^{abcd} \equiv f^{abc}{}_{e} h^{ed}$
to be totally anti-symmetric: 
\bea
 \label{tantis}
f^{abcd} = 
\frac{1}{4!}f^{[abcd]}  .
\eea
Remember that
(\ref{tantis}) is
guaranteed 
only for trace elements with invariant metric.
Inner product and metric are not defined 
for non-trace elements.
Nevertheless,
the 3-bracket can map non-trace elements 
to a trace element.
These non-trace elements will play important role in
this paper.
For more about Lie 3-algebra in the BLG model, see e.g.
\cite{Ho:2008bn,Papadopoulos:2008sk,Gauntlett:2008uf,%
Gomis:2008uv,Benvenuti:2008bt,Ho:2008ei,%
Lin:2008qp,%
FigueroaO'Farrill:2008zm,deMedeiros:2008bf}. 

The Bagger-Lambert action is given by
\cite{Bagger:2007jr}
%{Bagger:2006sk,Bagger:2007vi}and Gustavsson
%\cite{Gustavsson:2007vu,Gustavsson:2008dy} 
\bea
S = \int d^3 x \; {\cal L}, 
\eea
where the Lagrangian density ${\cal L}$ is given by
\bea
\label{BLaction}
&&{\cal L} = -\frac{1}{2} \la D^{\mu}X^I, D_{\mu} X^I\ra 
+ \frac{i}{2} \la\bar\Psi, \Gamma^{\mu}D_{\mu}\Psi\ra 
+\frac{i}{4} \la\bar\Psi, \Gamma_{IJ} [X^I, X^J, \Psi]\ra \nn \\ 
&&\qquad \quad -V(X) + {\cal L}_{CS}. 
\eea
$X^I \in {\cal V}_{tr}$\footnote{%
Later we will
relax this condition slightly and
allow constant backgrounds $X^I$ to
take values in non-trace elements.} %= \sum_a X^I_a T^a$ %($I = 1,\cdots,8$) 
is a scalar field on the worldvolume
and $I$ is a $SO(8)$ vector index. 
% we restrict the sum $\sum_a$
%for trace elements.
%In the later section, we will
%relax the above condition slightly and
%allow constant backgrounds of $X^I$ to
%take values in non-trace elements.
$\Psi \in {\cal V}_{tr}$ %= \sum_a \Psi_a T^a$
 are Majorana spinors on $1+2$ dimensional worldvolume,
but can be combined into a single Majorana spinor in eleven dimensions
subject to the chirality condition 
$\Gamma \Psi = - \Psi$, $\Gamma \equiv \Gamma_{012}$.
Notations for gamma matrices are summarized in the appendix.
%$A_\mu$ is a worldvolume gauge field and
$D_{\mu}$ is the covariant derivative 
\be
 \label{Dmu}
(D_\mu  X^I(x))_a = \partial _{\mu} X^I_a(x) -\tilde{A}_\mu{}^b{}_a(x) X^I_b(x), 
\quad 
\tilde{A}_{\mu}{}^b{}_a \equiv A_{\mu cd} f^{cdb}{}_a ,
\ee
where $A_\mu$ is a worldvolume gauge field.
$V(X)$ is the potential %term %defined by 
\be
V(X) = \frac{1}{12}\la [X^I, X^J, X^K], [X^I, X^J, X^K]\ra .
\ee
The Chern-Simons term for the gauge potential is given by
\bea
\label{CS}
{\cal L}_{CS} = \frac{1}{2}\ve^{\mu\nu\lambda}
\left(f^{abcd}A_{\mu ab}\pa_{\nu}A_{\lambda cd} 
+ \frac{2}{3} f^{cda}{}_g f^{efgb} A_{\mu ab} A_{\nu cd} A_{\lambda ef} \right). 
\eea
The Bagger-Lambert action is invariant under
the following gauge transformation:
\bea
 \label{gauge}
\delta_\Lambda X^I_a &=& 
\Lambda_{cd}[T^c,T^d,X^I]_a =\Lambda_{cd} f^{cde}{}_a X^I_e
=\tilde{\Lambda}^e{}_a  X^I_e, \nn\\
\delta_\Lambda \Psi_a &=&  
\Lambda_{cd}[T^c,T^d,\Psi]_a 
= \Lambda_{cd} f^{cde}{}_a \Psi_e
= \tilde{\Lambda}^e{}_a  \Psi_e, \nn\\
\delta_\Lambda \tilde{A}_{\mu}{}^b{}_a 
&=&
\pa_\mu  \tilde{\Lambda}_{\mu}{}^b{}_a 
-\tilde{\Lambda}^b{}_{c} \tilde{A}_{\mu}{}^c{}_a
+ \tilde{A}_{\mu}{}^b{}_c  \tilde{\Lambda}^c{}_{a},
\quad \tilde{\Lambda}^b{}_{a} 
\equiv f^{cdb}{}_a \Lambda_{cd}. 
%\delta_\Lambda A_{\mu ab} 
%&=&
%pa_\mu \Lambda_{ab} 
%-  f^{cde}{}_a A_{\mu cd} \Lambda_{eb} 
%+ f^{cde}{}_b A_{\mu cd} \Lambda_{ea}
%&=& (D_\mu \Lambda)_{ab},
\eea
%where
%\bea
%(D_\mu \Lambda)_{ab} &\equiv&
%\pa_\mu \Lambda_{ab} 
%-  f^{cde}{}_a A_{\mu cd} \Lambda_{eb} 
%+ f^{cde}{}_b A_{\mu cd} \Lambda_{ea} \nn \\ 
%&=& \pa_\mu \Lambda_{ab} 
%- \tilde{A}^e{}_a \Lambda_{eb}
%+ \tilde{A}^e{}_b \Lambda_{ea} .
%\eea
The fundamental identity (\ref{FI}) 
leads to
%for any $Y = \sum_a Y_a T^a$
%which transforms as
%$\delta_\Lambda Y = \Lambda_{ab}[T^a,T^b,Y]$,
\bea
 \label{tr3bra}
 \delta_\Lambda [\Phi(1),\Phi(2),\Phi(3)] =
% \tilde{\Lambda}^b{}_a [Y_1,Y_3,Y_3]_b,
 \Lambda_{cd}[T^c,T^d,[\Phi(1),\Phi(2),\Phi(3)]],
% \quad \tilde{\Lambda}^b{}_a \equiv f^{cdb}{}_a \Lambda_{cd}.
\eea
where $\Phi$'s collectively represent $X^I$ and $\Psi$.
The metric is not involved in reaching (\ref{tr3bra}) 
and this formula applies to both trace elements and non-trace elements.
On the other hand, 
the invariance of the metric (\ref{invm}) leads to
\bea
 \label{invm2}
 \delta_{\Lambda}
\la Y , Z \ra
=
\Lambda_{ab} 
\left(
\la
[T^a,T^b,Y],Z
\ra
+
\la
Y, [T^a,T^b,Z]
\ra
\right)
=0 .
\eea
for any trace elements $Y,Z$
%$Y = \sum_a Y_{a} T^a$, $Z = \sum_a Z_{a} T^a$
which transform as
$\delta_\Lambda Y
= \Lambda_{cd}[T^c,T^d,Y]$,
$\delta_\Lambda Z
= \Lambda_{cd}[T^c,T^d,Z]$.
(\ref{tr3bra}) and (\ref{invm2})
can be used to show
the gauge invariance of 
the Bagger-Lambert action.
%for any $Y=\sum_a Y_a T^a$
%which transforms as
%$\delta_\Lambda Y = \Lambda_{ab}[T^a,T^b,Y]$.

\subsection{Worldvolume supersymmetry of the BLG model}

The Bagger-Lambert action
%(\ref{BLaction}) 
is invariant under the following
supersymmetry transformations:\footnote{%
See \cite{Mauri:2008ai} for a ${\cal N}=1$ superfield formalism.}
\bea
 \label{WVSUSY}
\delta_\e X^I_a &=& i\bar{\epsilon}\Gamma^I \Psi_a, \nn \\
\delta_\e \Psi_a &=& D_{\mu}X^I_a \Gamma^\mu\Gamma^I \epsilon  
- \frac{1}{6} X^I_b X^J_c X^K_d f^{bcd}{}_a \Gamma^{IJK}\epsilon, \nn \\
\delta_\e \tilde{A}_{\mu}{}^b{}_a &=& 
i\bar{\epsilon}\Gamma_{\mu}\Gamma_I X^I_c \Psi_d f^{cdb}{}_a, 
\eea
where the supersymmetry parameter
satisfies $\Gamma \epsilon = \epsilon$.
The charge density, i.e. 
the temporal component of the Noether current 
associated with the supersymmetry transformation
(\ref{WVSUSY}), is found to be
\bea
 \label{ql}
\ql 
=
-
\Gamma^\mu\Gamma^I\Gamma^0
\la
D_\mu X^I,
\Psi
\ra
-
\frac{1}{6}
\Gamma^{IJK}
\Gamma^0
\la
[X^I,X^J,X^K],
\Psi
\ra ,
\eea
and the Noether charge is 
\bea
 \label{QL}
\QL = \int d^2x \, \ql  .
\eea
The suffix $L$ indicates that it is identified
with the linearly realized part of the space-time
supersymmetry.

In this paper we will often be interested
in central charges which are proportional to
the volume of the membranes, which can be
infinite for infinitely extended membranes.
A standard way to avoid infinities
associated with such infinite volume
arising from the (anti-)commutation relations of 
Noether charges
is to use charge density.
In the following, it is understood that
the fermions $\Psi$ are set to zero after
calculating the Dirac bracket,
since we are interested in bosonic backgrounds.
The Dirac bracket of
$\ql$ and $\QL$
is calculated to be
\bea
\label{QLQL}
i
\{\ql, \QL \}_D &=&
2 p_\mu  \Gamma_+ \Gamma^\mu C  \nn \\
&+&
z_{IJ} \Gamma^{IJ} C 
+
z_{0ijIJ}  \Gamma^{0ijIJ} C \nn \\ %\label{QLQL2}\\
&+&
z_{0iIJKL} \Gamma^{0iIJKL} C
+
z_{jIJKL} \Gamma^{jIJKL} C  \nn \\ %\label{QLQL3}
&+&
z_{0IJKL} \Gamma^{0IJKL} C
+
z_{ijIJKL} \Gamma^{ijIJKL} C,
\eea 
where 
\bea
 \label{zIJ}
z_{IJ} 
=
\frac{1}{2}
\left(
-
\ve^{0ij}
\la
 D_i X^I,D_j X^J
\ra
+
\la
D_0 X^K ,
[X^K,X^I,X^J]
\ra
\right) ,
\eea
\bea
 \label{z0ijIJ}
z_{0ijIJ} 
=
\frac{1}{2}
\left(
\la
 D_i X^I,D_j X^J
\ra
-
\frac{1}{2}\ve^{0}{}_{ij}
\la
D_0 X^K ,
[X^K,X^I,X^J]
\ra
\right) ,
\eea
\bea
 \label{z0iIJKL}
z_{0iIJKL}
=
\frac{1}{6}
\la
D_i X^I ,
[X^J,X^K,X^L]
\ra ,
\eea
\bea
 \label{ziIJKL}
z_{iIJKL} =
-
\frac{1}{6}
\ve^{0j}{}_{i}
\la
D_j X^I,[X^J,X^K,X^L] 
\ra  ,
\eea
\bea
 \label{z0IJKL}
z_{0IJKL}
=
-\frac{1}{8}
\la
[X^M,X^I,X^J],[X^M,X^K,X^L]
\ra ,
\eea
\bea
 \label{zijIJKL}
z_{ijIJKL}
=
-\frac{1}{16}
\ve^{0}{}_{ij}
 \la
[X^M,X^I,X^J],[X^M,X^K,X^L] 
\ra .
\eea
In the above, %$\Gamma_\pm \equiv (1 \pm \Gamma)/2$, and
anti-symmetrization of the space-time indices %$i,j$ and $I,J,K,L$ 
is understood. 
And $\Gamma_\pm \equiv (1 \pm \Gamma)/2$.
This projection arises from the chirality of the supercharges:
$\Gamma \QL = \QL$.
In the second, the third and the fourth lines of (\ref{QLQL}), 
two terms in the same line 
arise from two different Gamma matrices
in the projection $\Gamma_+ = (1+\Gamma)/2$.
The bosonic part of the Hamiltonian density is given by
\bea
{\cal H} = p_0  = 
\frac{1}{2} \la D_0 X^I , D_0 X^I \ra
+
\frac{1}{2}
\la D_i X^I , D_i X^I \ra
+
V(X) ,
\eea
and the momentum density is given by
\bea 
p_i = \la D_0 X^I, D_i X^I \ra  .
\eea
%Some calculational steps are collected in the appendix.
We refer to the appendix for details.
These central charges have been discussed
in \cite{Passerini:2008qt};\footnote{%
The expressions for the central charges
look different 
just because
we haven't used the
equation of motions.}
the combination of 
the central charges
(\ref{zIJ}) and (\ref{z0ijIJ})
was found to be the charge of 
vortices, and identified with
M2-branes intersecting with the 
multiple M2-branes.
The combination of 
the central charges
(\ref{z0iIJKL}) and (\ref{ziIJKL})
was found to be the charge of 
Basu-Harvey solution
\cite{Basu:2004ed}
which had been identified with
M2-branes ending on M5-branes.
Readers interested in further discussions
are advised to consult \cite{Passerini:2008qt}.

%(the expressions for the central charges
%look different but this is 
%just because
%we haven't used
%equation of motions).
%Since those charges have already been 
%discussed in \cite{Passerini:2008qt},
%we will not explain them in detail in this paper.

The central charges (\ref{z0IJKL}) and (\ref{zijIJKL})
vanish if we only consider 
trace elements in the Lie 3-algebra %3-brackets
%which have an invariant inner product,
due to the total anti-symmetry of the indices
$I,J,K,L$ and the fundamental identity (\ref{FI}).
However,
we would like to 
take into account
%the worldvolume coordinates
%independent
constant background configurations of $X^I$'s which
%independent of the worldvolume coordinates
%where some $X^I$'s
take values in non-trace elements.
As long as they give trace elements
after putting into the 3-brackets in
the Bagger-Lambert action, 
%it is easy to convince oneself 
%(see (\ref{tr3bra}) and (\ref{invm2}))that
the action is still well-defined
and gauge invariant,
provided the fluctuations around the background
are still restricted to trace elements.\footnote{%
Recall (\ref{tr3bra}) and (\ref{invm2}).
The configuration is gauge covariant,
but the value of the action is invariant under
gauge transformations with parameters
taking values in trace elements.}
This kind of configurations give
rise to BPS brane charges.
For example, 
in the case when the
Bagger-Lambert action reduces to D2-brane action
by giving expectation value to the field
$X^{10}_{0}$ in the notation of
\cite{Ho:2008ei},
(\ref{z0IJKL}) and (\ref{zijIJKL})
reduce to the form
$
\sim \mbox{tr} [X^I,X^J][X^K,X^L]
$,
where $[*,*]$ is the commutator of matrices
and tr is the trace for matrices, and the matrix size
is to be taken to infinity.\footnote{%
In this case, actually the commutator of $X^I$'s are
still non-trace elements, and the central charge diverges.
This is attributed to the infinite volume of
indefinitely extended D6-branes
discussed below.
The charge density per D6-brane worldvolume is still finite.}
% in finite D6-brane volume.}
%The D2-brane charge within D6-branes,
%the analogue of the D0-brane charge within D4-branes, is finite.}
This term is analogous to the D4-brane charge
(as well as the charges of the D0-branes within the D4-branes)
in the matrix model for M-theory found 
in \cite{Banks:1996nn}, and one should keep
this kind of terms in order to obtain all
the BPS-brane charges in the model.
In the current example, the action reduces to D2-branes
instead of D0-branes for the matrix model,
so the charge should be interpreted as D6-brane charge.
This type of configuration
is also crucial for the construction
of the five-brane from the BLG model in \cite{Ho:2008nn,Ho:2008ve}
and we will discuss this in section \ref{M5}.

\subsection{Space-time superalgebra from the BLG model}

It has been noticed that the
choice of Lie 3-algebra
in the BLG model already contains the
choice of space-time in which membranes are embedded
\cite{VanRaamsdonk:2008ft,Lambert:2008et,Distler:2008mk}.
This is not surprising if we recall 
the analogous situation in
multiple D-brane worldvolume theory,
where the gauge group
contains information of space-time,
e.g. orientifold for gauge group $SO$.
%or geometric realization of quantum field theories on D-branes.
In the BLG model,
when there is a central element in the Lie 3-algebra
there is a bosonic shift symmetry in this direction: 
%which can be identified with
%the translation transverse to membranes:
\bea
 \label{trans}
&&\quad \delta_{\vec{a}} X^I_\C = a^I \quad (a^I : \mbox{constant}), \nn \\
&&\delta_{\vec{a}} \Psi_a = \delta_{\vec{a}} \tilde{A}_\mu{}^b{}_a=0,
\eea
as well as the fermionic shift symmetry\footnote{%
The fermionic shift symmetry has been used
in \cite{Ho:2008ve}
to obtain the worldvolume 
supersymmetry of the five-brane action
constructed from the BLG model.}
\bea
 \label{shift}
&&\quad {\delta}_\eta \Psi_a = \delta_{a\C} \eta, \nn \\
&&{\delta}_\eta X^I_a = \delta_\eta \tilde{A}_\mu{}^b{}_a=0 .
\eea
We use index $\C$ to denote the generator 
corresponding to the central element. 
In the following, we 
restrict ourselves to the case
where the metric for this central element takes
following form:
%where there is only one central element in the algebra.
%We further assume that the inner product is positive definite
%and the metric can be taken as 
\bea
 \label{cmet}
%h^{\C \C} &=& 1 , \nn \\
h^{\C a} &=& \delta^{\C a}.
\eea
%and the generator $T^\C$ 
%is one of the ortho-normalized basis.
With this metric, it is natural to identify
$X^I_\C$ as 
the center of mass coordinate in
the direction transverse to membranes up to normalization,
and (\ref{trans}) is nothing but the 
translational symmetry in this direction.
We further assume that there is only one such
central element with metric of the form
(\ref{cmet}) in the Lie 3-algebra,\footnote{%
We allow other central elements with non-positive-definite
metric \cite{Gomis:2008uv,Benvenuti:2008bt,Ho:2008ei}.}
%The constraints from requiring
%the fundamental identity (\ref{FI}) is very restrictive
%and in this case it has been shown that
%the Lie 3-algebra must be
%direct sum of trivial algebra or
%so-called ${\cal A}_4$ algebra or
%infinite dimensional algebra
%\cite{Papadopoulos:2008sk,Gauntlett:2008uf}.
%A Lie 3-algebra with non-positive-definite metric
%has been studied in
%\cite{Gomis:2008uv,Benvenuti:2008bt,Ho:2008ei}, 
%but this
%led to the action of multiple D2-branes.
%We are more interested in the structure
%genuinely associated with the BLG model or M-theory
%and this is the reason for the above choice,
%although interesting examples of Lie 3-algebra
%with non-positive definite metric
%may be found in the future.
%We hope the essential part of our arguments 
%may go through without 
%significant changes even in such cases.
%This is 
because %we would like to identify 
%if $X^I_\C$
%is the center of mass coordinate,
it is strange
if there are 
two sets of center of mass coordinates.\footnote{%
Though it may work with some kind of gauging.}
In our setting,
the Noether charge density associated 
with the transformation (\ref{shift}) is
given by
\bea
 \label{tq}
\tq = - \Gamma^0 {\Psi}_\C  ,
\eea
and the Noether charge is 
\bea
 \label{tQ}
\tQ = \int d^2x \, \tq ,
\eea
where the suffix $NL$ indicates that 
it is identified with the non-linearly realized part
of the space-time supersymmetry. 
Note that $\tQ$ has the same chirality with 
the worldvolume fermions $\Psi$,
i.e. $\Gamma \tQ = - \tQ$, as opposed to $\QL$.

The BLG model is not space-time
super-Poincar\'{e} symmetric,
at least manifestly.
However, if we want to regard the model as
a description of multiple M2-branes,
it is natural to expect that 
it is a gauge fixed form of some
space-time supersymmetric and
worldvolume reparametrization invariant formulation.
In the case of single supermembrane,
it has been shown that the space-time supersymmetry
reduces to the worldvolume supersymmetry
by static gauge fixing 
%and a suitable gauge choice of $\kappa$ symmetry
\cite{Achucarro:1987nc,Bergshoeff:1987qx,Achucarro:1988qb}.
After the gauge fixing, 
the linearly realized part of the space-time supersymmetry
becomes global supersymmetry on the worldvolume theory, 
whereas
the Nambu-Goldstone modes for the
non-linearly realized part of the space-time
supersymmetry become fermion fields on the worldvolume \cite{Hughes:1986dn}.
In our case, fields
$\Psi$ can be thought of as Nambu-Goldstone fermions
for non-linearly realized space-time supersymmetry.
In the following we will show that the charge $\tQ$ 
associated with the fermionic shift symmetry (\ref{shift})
almost provides 
the non-linearly realized part of the space-time supersymmetry,
%(the suffices $L$ and $NL$ 
%in (\ref{ql}), (\ref{QL}) and (\ref{tq}), (\ref{tQ})
%indicate that they are the linearly and 
%%(a part of) 
%the non-linearly
%realized supersymmetry, respectively),
though there is a missing piece 
as we will see shortly.
%(see (\ref{QN})).

The Dirac bracket of
%commutation relation between 
$\tq$ and $\QL$ are given as
\bea
 \label{QNLQL}
&&i 
\{
\tq, \QL
\}_D 
+
i 
\{
\ql, \tQ
\}_D \nn \\
&=&   p_I  \Gamma^I C 
+ 
\frac{1}{2} %due to the \tq's normalization
z_{iI} \Gamma^{iI} C 
+ 
\frac{1}{2} %due to the \tq's normalization
z_{ijIJK} \Gamma^{ijIJK} C  ,
\eea
where
\bea
p_I \equiv \pa_0 X^I_\C
\eea
is the momentum density
in the direction transverse to the membranes.
%Noether charge density for the symmetry (\ref{trans})
%which is interpreted as the space-time translation symmetry
%in the direction transverse to M2-branes.
The central charge densities are found to be
\bea
 \label{tilt}
z_{iI}
=
2 %due to the \tq normalization
 \pa_j  X^I_\C \ve_i{}^{j0} \, ,
\eea
\bea
\label{5charge}
z_{ijIJK}
=
-
\frac{1}{6} %due to the \tq normalization
\varepsilon^0{}_{ij}
\langle
[X^I,X^J,X^K ],T_\C
\rangle .
\eea
The central charge density (\ref{tilt}) 
describes tilting multiple membranes.
%expresses the charge for a
%tilt of the multiple membranes.
For example, let us consider the situation where
$X^I_\C$ is compactified on a circle with radius $R^I$,
and $x^j$ is compactified on a circle with radius $r^j$.
Then %the configuration
\bea
X^I_\C = \frac{n}{m} \frac{R^I}{r^j} x^j
\eea 
is a configuration of membranes 
which winds the $I$-th direction for $n$
times and $j$-th direction for $m$ times.
This configuration 
gives topological winding numbers through
the central charge density (\ref{tilt}).

The central charge density (\ref{5charge}) vanishes
when all $X^{I}$'s in the 3-bracket 
take values in trace elements of the Lie 3-algebra,
%which have an invariant metric,
due to the 
definition of the central element and 
the invariance of the metric (\ref{invm}).
%total anti-symmetry of the indices of the
%structure constants (\ref{tantis}).
This is not necessarily the case
if we consider %take into account 
constant configurations
where $X^I$'s take values in non-trace elements.
%However, recall that the anti-symmetry property (\ref{tantis})
%was due to
%the existence of the invariant inner product (\ref{invm}).
%If we regard the 3-bracket $[\a,\b,*]$
%as an analogue of
%derivation in ordinary algebra,
%there can be an analogue of
%surface term or topological winding
%for elements of Lie 3-algebra
%which do not have an invariant inner product.
%This can happen when we deal with 3-algebras
%with infinitely many generators.
As long as we obtain a trace element
after putting such $X^I$'s 
into the 3-bracket,
the inner product in (\ref{5charge})
is well-defined and gives a finite number.
The Bagger-Lambert action is also well-defined
for such configuration, provided it is regarded
as a background;\footnote{%
Note that the covariant derivative (\ref{Dmu})
can be rewritten as
$D_\mu X^I = \pa_\mu X^I - A_{\mu\,cd}
[T^c,T^d,X^I]$.}
fluctuations from the background should still
be in the space of trace elements.
To describe a five-brane
%extending in $R^3$ or wrapping on $T^3$ 
in the BLG model
based on Nambu-Poisson bracket \cite{Ho:2008nn,Ho:2008ve},
the background configuration 
is indeed given by such $X^I$'s taking values in non-trace elements, 
and (\ref{5charge})
gives the charge of the five-brane.
We will come back to this point again in
section \ref{M5}.

The Dirac bracket of $\tq$ and $\tQ$ is given by
\bea
 \label{tQtQ}
i \{ \tq, \tQ \}_D
= 
%\frac{1-\Gamma}{2}
\Gamma_- \Gamma^0 C 
=
\frac{1-\Gamma}{2}
\Gamma^0 C  .
%\dim {\cal A}.
\eea
%where $\Gamma_\pm \equiv (1\pm \Gamma)/2$.
The last expression in (\ref{tQtQ}) can be 
interpreted as a sum of
%space-time energy density
the mass density
and the charge density
of the static multiple membranes.
(The absence of such term in (\ref{QLQL})
can be regarded as cancellation of mass and charge
for the BPS configuration of membranes \cite{Witten:1978mh}.)
However, it does not contain 
contributions from excitations 
on the worldvolume
to the energy nor 
the momentum in the worldvolume direction,
which are required for making up the eleven dimensional
super-Poincar\'e algebra.
Besides this point,
we can construct full space-time supercharge density $q$ and 
charge $Q$ as follows:
\bea
q = \ql + 2 \sqrt{N}
\tq,
\eea
\bea
Q = \QL + 2 \sqrt{N} \tQ .
\eea
Here, we have introduced a constant
$N$ which
can be interpreted as ``number" of membranes.
The reason for this factor is as follows:
Up to normalization $X^I_\C$ is interpreted as 
the center of mass coordinate in the transverse directions.
To define the center of mass, we need to know
the number of membranes.
However, there's no definite rule for relating
the dimension of a Lie 3-algebra
and the number of membranes.
In the case of the
Lie 3-algebra constructed from
ordinary Lie algebra in order
to derive D2-brane action from the Bagger-Lambert action
\cite{Gomis:2008uv,Benvenuti:2008bt,Ho:2008ei},
the number of membranes
should be equal to the number of D2-branes and
determined from the rank of the Lie group,
e.g. $N$ for $U(N)$.
We will discuss the case when
Nambu-Poisson bracket is chosen
as Lie 3-algebra in the next subsection.
Since 
%there's no definite rule for
%relating the dimension of Lie 3-algebra and
the number of membranes is
decided case by case depending on 
the choice of Lie 3-algebra, 
we just denote
this number as $N$.

Finally, we obtain the eleven dimensional super-Poincar\'e algebra
with central extensions:
\bea
 \label{QQ}
i \{q,Q\}_D
&=&
2 (\Gamma^0 - \Gamma^{12})  C  N
 + 2 p_\mu \Gamma_+ \Gamma^\mu C 
 + 2 p_I \Gamma^I C  \sqrt{N} \nn \\
&+& 
  z_{iI} \Gamma^{iI} C \sqrt{N} 
+ z_{ijIJK} \Gamma^{ijIJK} C \sqrt{N} 
\nn \\
&+& 
z_{IJ} \Gamma^{IJ} C 
+
z_{0ijIJ}  \Gamma^{0ijIJ} C \nn \\
&+&
z_{0iIJKL} \Gamma^{0iIJKL} C
+
z_{jIJKL} \Gamma^{jIJKL} C  \nn \\
&+&
z_{0IJKL} \Gamma^{0IJKL} C
+
z_{ijIJKL} \Gamma^{ijIJKL} C.
\eea
As mentioned above,
the first term of (\ref{QQ}) is interpreted 
as coming form tension and charge per volume
of $N$ membranes.
From the kinetic term the relative normalization between 
$X^I_\C$ and the center of mass coordinate
is read off as
$X^I_\C = \sqrt{N} X^I_{C.M.}$,
where $X^I_{C.M.}$ is the center of mass coordinate.
Then $p_I \sqrt{N} = p_{I}^{C.M.} N$ 
is the total momentum 
in the direction transverse to membranes,
and $N$ appears in an appropriate way
for a number of membranes.

Eq. (\ref{QQ}) is almost the 
eleven dimensional super-Poincar\'e algebra,
except that the piece 
$2p_\mu \Gamma_- \Gamma^\mu C$ is missing
in the right hand side of (\ref{QQ}).
It is important that the piece  
$2p_I \Gamma^I C$ 
for the eleven dimensional super-Poincar\'e algebra
has been obtained.
If we had a
space-time supersymmetric formulation 
with worldvolume reparametrization invariance
for
multiple membrane action which reduces
to the Bagger-Lambert action after gauge fixing,
this would be understood as due to 
the static gauge 
and kappa symmetry gauge fixing.
We hope to clarify this point in the future.
Further speculations will be given
in the last discussion section.

\subsection{On M5-brane charges in the BLG model}
\label{M5}

An example of Lie 3-algebra is given by
Nambu-Poisson bracket on an ``internal" three-manifold.
For simplicity,
we take $T^3$ to be the internal three-manifold.
For more about the use of
Nambu-Poisson bracket in the BLG model, see
\cite{Ho:2008bn,Ho:2008nn,Ho:2008ve}.
We consider the Nambu-Poisson bracket on $T^3$ given by
\bea
 \label{Nambu}
\left\{f,g,h\right\}_{\rm NP}
=
\sum_{\dot\mu\dot\nu\dot\lambda}
\,  \Omega  \ve^{\dot\mu\dot\nu\dot\lambda}\partial_{\dot\mu}
f(y) \partial_{\dot\nu} g(y) \partial_{\dot\lambda} h(y)  .
\eea 
Here $y^{\dot\mu}$ ($\dot\mu=\dot 1,\dot2,\dot3$) 
are flat coordinates on $T^3$ with the identification
$y^{\dot\mu} \sim y^{\dot\mu}+ 2\pi$, and
$\Omega$ is a constant.
The invariant inner product can be defined by the integral
over  $T^3$:
\begin{equation}
\langle f,g\rangle
\equiv
\int_{T^3} d^3y \, f(y) g(y).
\end{equation}
The trace elements of the Lie 3-algebra
%with invariant inner products
are given by %square integrable smooth (infinitely differentiable)
square-integrable periodic functions on $T^3$.
If we denote the basis of such %square integrable periodic
functions on $T^3$ as
$\chi^a(y)$ ($a=1,2,3,\cdots$), the Nambu-Poisson bracket
can be written with structure constants: %as a Lie 3-algebra
\bea
\{ \chi^a, \chi^b, \chi^c\}_{\rm NP}=\sum_d {f^{abc}}_d \chi^d\, .
\eea
Using the definition of the 
Nambu-Poisson bracket (\ref{Nambu}),
it is
easy to check that the fundamental identity (\ref{FI}) holds.
We normalize the basis as $\la \chi^a ,\chi^b \ra = \delta^{ab}$;
then the normalized central element is given as
$T^\C = 1/\sqrt{(2\pi)^3}$.

We would like to consider the case where the target space
is also compactified on a $T^3$.
By this we mean the identification in the central element:
\bea
 \label{ttorus}
X^I(y) \sim X^I(y) + 2\pi R^I ,
\eea
for say $I=3,4,5$,
where $R^I$ is the compactification radius in
the $I$-th direction. 

Now let us consider a
background configuration 
\be
 \label{wrap}
X^{I}=  R^I m_{I} y^{\dot{\mu}},\quad \dot\mu = I-2\quad (I = 3,4,5).
\ee
The functions $y^{\dot\mu}$ 
($\dot\mu = \dot{1}, \dot{2}, \dot{3}$)
are not periodic functions on $T^3$:
%included
%in the square integrable functions on $T^3$
%because they do not satisfy
%periodic boundary conditions on $T^3$:
They have a jump at $y^{\dot\mu} = 2\pi$.
%We cannot define invariant inner product for this element.
However, when the target space is
also compactified as in (\ref{ttorus}), 
such jump can be set to null %equivalent to zero 
for the configuration
(\ref{wrap}) due to the identification in the target space.
In this case, it is natural to include these elements
in the Lie 3-algebra.
However, there is no natural way to define
invariant inner product for these elements.
For example,
\bea
&&\int_{T^3} d^3y \, \{ y^{\dot{1}} , y^{\dot{2}} , 1 \}_{NP} \cdot y^{\dot{3}} =0 \nn \\
&\ne& - \int_{T^3} d^3y \, 1 \cdot \{ y^{\dot{1}} , y^{\dot{2}} , y^{\dot{3}} \}_{NP} 
= \Omega (2\pi)^3  .
\eea
This means that the integration over $T^3$ does not
provide an invariant metric for these new elements.
Therefore, these elements should be included
as non-trace elements.
In the Bagger-Lambert action,
these $X^I$'s in non-trace elements
%configuration
always appear inside the
Nambu-Poisson brackets;
and the Nambu-Poisson brackets
with such non-trace elements
give trace elements,
since the derivative inside the Nambu-Poisson bracket
acting on $y^{\dot\mu}$ 
gives a constant which is a trace element.
As long as such configuration is regarded as 
a non-dynamical background
independent of the worldvolume coordinates,
%i.e.
%we do not allow fluctuations to be non-trace elements,
the Bagger-Lambert action
is still well-defined and gauge invariant.
%\footnote{%
%The background configuration is gauge covariant,
%but the value of the action is invariant under
%gauge transformations with parameters
%taking values in trace elements.}

Now we come back to the issue of the ``number of membranes"
discussed in the previous subsection.
In the current case
where there is a natural notion of identity ``1"
in the elements and the metric is positive definite, %it turns out that
it is natural to interpret 
the number we get when we put ``1" into the inner product
as the number of membranes.
This is nothing but the volume of the internal manifold $T^3$.
Therefore we set $N = (2\pi)^3$.

The background configuration (\ref{wrap}) contributes to the
five-brane charge (\ref{5charge}) as
\bea
 \label{ex5}
z_{ijIJK} \sqrt{N}
=
-
\frac{1}{3!} \ve_{IJK}
{(2\pi)^3} % due to the \tq normalization
\ve^0{}_{ij}
\Omega R^3 R^4 R^5
m_{3}m_{4}m_{5}  .
\eea
%\cite{Ho:2008nn,Ho:2008ve}.
(\ref{ex5})
is interpreted as a charge of a five-brane
wrapping the $I$-th direction for $m_I$ times.

Note that the potential term in the Bagger-Lambert action
can be rewritten as
\bea
&&V(X) \nn \\
&=&
\frac{1}{12}
\biggl(
\la
[X^I,X^J,X^K] - W^{IJK} T^\C,
[X^I,X^J,X^K] - W^{IJK} T^\C
\ra
\nn \\
&&\qquad
+ 
2
W^{IJK}
\la
[X^I,X^J,X^K], T^\C
\ra
-
W^{IJK}W^{IJK}
\biggr)\nn \\
&=&
\frac{1}{12}
%\biggl(
\la
[X^I,X^J,X^K] - W^{IJK} T^\C,
[X^I,X^J,X^K] - W^{IJK} T^\C
\ra
\nn \\
&&\qquad
-
\frac{1}{2} 
W^{IJK}
\ve^{0ij}
z_{ijIJK}
-
\frac{1}{12}
W^{IJK} W^{IJK} ,
%\biggr),
\eea
where $W^{IJK}$ is a constant
totally anti-symmetric tensor 
\bea
W^{IJK} = \ve^{IJK}\Omega R^3R^4R^5m_3m_4m_5  .
\eea
Therefore the static configuration (\ref{wrap})
saturates the minimal energy bound
for given winding numbers.

Some time ago 
a matrix model was proposed as a description of
M-theory \cite{Banks:1996vh}, and
BPS branes in this model were analyzed
from the central extension of the superalgebra \cite{Banks:1996nn}.
It was found that the charge of
transverse five-branes, i.e. five-branes transverse
to the M-theory circle which relates %compactified to relate
M-theory to type IIA string, is absent in this model.
This can be a problem
if the model is the fundamental definition of 
M-theory, %which should contain all the brane charges,
though the model may better be regarded
as M-theory in a particular frame
%, namely infinite momentum frame, 
in which some information of the
full M-theory has been dropped off.
From our results for the M-theory superalgebra (\ref{QQ}),
we can draw 
a scenario for how such thing can happen in the BLG model:
The action for the matrix model for M-theory 
is basically that of the
large number of multiple D0-branes in type IIA string theory.
From the Bagger-Lambert action, 
such action may be obtained by 
first reducing it to multiple D2-brane action
\cite{Mukhi:2008ux,Gomis:2008uv,Benvenuti:2008bt,Ho:2008ei},
then wrapping D2-branes on $T^2$, 
and then performing T-duality transformations
in the $T^2$ directions.
To obtain
multiple D2-brane action from the Bagger-Lambert action,
it is necessary to reduce Lie 3-algebra 
to ordinary Lie algebra.
%This was achieved by 
%giving an expectation value to one of $X^I$, say $X^{10}$ \cite{Mukhi:2008ux}.
%This direction is identified with the direction of the
%circle through which
This should be achieved by some background configuration
in the BLG model which describes a compactification 
on the M-theory circle.
%through which
%M-theory is related to type IIA string theory. 
However, by this background configuration
the five-brane charges
expressed using Lie 3-algebra in 
(\ref{z0iIJKL}), (\ref{ziIJKL}) or (\ref{5charge})
must also reduce to the expression
using ordinary Lie algebra.
This should be interpreted as 
five-branes %in this model
are also wrapping the circle direction.
%because such ordinary Lie algebra
%should be the world volume gauge group in D4-branes.
Thus when one obtains
the matrix model for M-theory from the Bagger-Lambert action,
transverse five-brane charges
which uses Lie 3-algebra structure
in an essential way, 
i.e. those which do not reduce to a form
written with ordinary Lie algebra,
necessarily 
drop out from the model.

\section{Summary and discussions}

In this paper we studied 
the space-time supersymmetry of the BLG model
when there is a central element in the Lie 3-algebra,
and obtained the %central extension of 
eleven dimensional super-Poincar\'e
algebra with central extensions,
%or M-theory superalgebra,
except the piece $2 p_\mu \Gamma_- \Gamma^\mu C$.
The first crucial ingredient
in the construction of the space-time superalgebra
%the eleven dimensional 
%super-Poincar\'e algebra
was to
include the fermionic shift symmetry 
associated with the central element in the Lie 3-algebra.
This fermionic shift symmetry was identified
with the non-linearly realized part of the
space-time supersymmetry.
Together with the linearly realized worldvolume
supersymmetry, it makes up the
eleven dimensional super-Poincar\'e algebra.
The second important ingredient
was to take into account the
non-trace elements for constant background configurations.
The central charges constructed from
non-trace elements provide important pieces of the
M-theory superalgebra.
For example, the charge of the five-brane constructed in \cite{Ho:2008nn,Ho:2008ve}
can only be constructed by taking into account such
non-trace elements.

Compared with the matrix model for M-theory
which can be regarded as regularization of
supermembrane action in the light-cone gauge \cite{deWit:1988ig},
the BLG model lacks relation to a manifestly 
space-time supersymmetric formulation at this moment.
Nevertheless, in this paper we could obtain the eleven dimensional 
super-Poincar\'{e} algebra almost fully.
This suggests the existence of a
manifestly space-time supersymmetric formulation
with worldvolume reparametrization invariance
which reduces to the BLG model after gauge fixing.
It will be very interesting to 
construct such manifestly space-time supersymmetric
formulation for the BLG model,
and understand why the piece
$2 p_\mu \Gamma_- \Gamma^\mu C$ is missing
in our algebra.
In the case where the Lie 3-algebra is 
Nambu-Poisson bracket, it is likely that
such manifestly space-time supersymmetric formulation 
is some covariant formulation
of single M5-brane worldvolume action in three-form field background
rather than multiple M2-brane action:
If we can find a way to relate such formulation 
to the five-brane action
constructed from the Bagger-Lambert action in \cite{Ho:2008nn,Ho:2008ve},
we will be able to understand the origin of our super-Poincar\'e algebra.
An interesting worldvolume reparametrization invariant formulation
of single M5-brane action
which might be related to the BLG model
%along the above line of arguments
was constructed in \cite{Park:2008qe}, though only
the bosonic part has been worked out.
A worldvolume supergravity action which in a limit reduces
to the Bagger-Lambert action was constructed in
\cite{Bergshoeff:2008ix}.

When the Lie 3-algebra does not have a central element,
the fermionic shift symmetry is absent.
In this case the space-time supersymmetry should be less compared with the flat space.
This may be regarded as a supersymmetric counterpart of 
the absence of space-time translational symmetry 
in the orbifold interpretation
of the model based on so-called ${\cal A}_4$ algebra 
\cite{VanRaamsdonk:2008ft,Lambert:2008et,Distler:2008mk}.

The BLG model is
superconformal at the classical level,
and expected to be so at the quantum level.
The superconformal symmetry should correspond to the near horizon
super-isometry in AdS-CFT correspondence,
and this is one of the strongest motivations for
studying this model.
It will be interesting to construct central extension of
the superconformal algebra explicitly in the BLG model.

\section*{Acknowledgments}

We would like to express special thanks 
to Pei-Ming Ho for many helpful explanations and discussions.
We would also like to thank Yosuke Imamura
and Wen-Yu Wen for discussions, and
%We would also like to thank
Takayuki Hirayama 
and Dan Tomino for reading the 
manuscript and for useful comments.
This work is supported in part by 
National Science Council of Taiwan
under grant No. NSC 97-2119-M-002-001.

\appendix
\section*{Appendix}

\section{Notation for indices}
\label{indices}

\bea
\mbox{worldvolume coordinates}&:& \mu,\nu = 0,1,2 \nn \\
\mbox{spatial worldvolume coordinates}&:& i,j = 1,2 \nn \\
\mbox{transverse space coordinates}&:& I,J = 3, \cdots, 10 \nn \\
\mbox{all 11D coordinates}&:& m,n = 0,1, \cdots, 10 \nn \\
\mbox{$Spin(1,10)$ spinor indices}&:& \alpha,\beta = 1, \cdots, 32 \nn \\
\mbox{basis of Lie 3-algebra ${\cal A}$}&:& a,b, \cdots , \dim {\cal A}
\eea

\section{Eleven dimensional Clifford algebra}
\label{Gamma}

11D Gamma matrices
\bea
\{\Gamma^m, \Gamma^n \} = 2 \eta^{mn}
\quad (m,n = 0,1, \cdots, 10) .
\eea
We use mostly plus convention, i.e. 
$\eta_{00}=-1$,
$\eta_{mn}=\delta_{mn} (m,n \ne 0)$.
The charge conjugation matrix $C$ in 
eleven dimension satisfies
\bea
C^{-1} \Gamma^m C 
=
- (\Gamma^m)^T ,\quad
C^T = -C, \quad
C^{\dagger}C = 1 .
\eea
\bea
\Gamma^{m_1 m_2 \cdots m_r}
&\equiv&
\frac{1}{r !} \Gamma^{[m_1}\cdots \Gamma^{m_r]}\nn \\
&=& \Gamma^{m_1}\Gamma^{m_2} \cdots \Gamma^{m_r}\quad 
(\mbox{when all $m_s$ are different}). \nn\\
&=& 0 \quad (\mbox{otherwise}).
\eea
$\Gamma^{m_1m_2 \cdots m_r} C$ is
a symmetric matrix for $r=1,2,5,6,9,10$. 
%and anti-symmetric for $r=0,3,4,7,8$.
%Majorana condition for a spinor $\psi$ is 
%\bea
%\psi = C \bar{\psi}^T .
%\eea

\subsection{$Spin(1,2) \otimes Spin(8)$ decomposition}

We define
\bea
\Gamma \equiv \Gamma_{012} = 
\frac{1}{3!}
\ve^{\mu\nu\rho} \Gamma_\mu \Gamma_\nu \Gamma_\rho ,
\eea
where $\ve^{\mu\nu\rho}$ is the totally anti-symmetric tensor
with $\ve^{012} = 1$.
\bea
[\Gamma, \Gamma^\mu ] = 0, \quad \{ \Gamma, \Gamma^I \} = 0 .
\eea
\bea
C\Gamma^T = \Gamma C .
\eea
\bea
\Gamma_\pm \equiv 
\frac{1 \pm \Gamma}{2}  .
\eea
Decomposition:
\bea
\Gamma^\mu = \gamma^\mu \otimes \bar{\gamma}_9, \quad
\Gamma^I = 1 \otimes \bar{\gamma}^I,
\eea
where $\gamma^\mu$'s
are gamma matrices in (1+2)D and
$\bar\gamma$'s are that of 8D, and
\bea
\bar{\gamma}_9 \equiv \bar{\gamma}^3 \cdots \bar{\gamma}^{10}  .
\eea
If we choose the basis for the (1+2)D
Clifford algebra as
\bea
\gamma_0 = i\s_2, \quad \gamma_1 = \s_1, \quad \gamma_2 = \s_3 ,
\eea
then
\bea
\Gamma \equiv \Gamma_{012} = 1 \otimes \bar{\gamma}_9 ,
\eea
i.e. the chirality for $\Gamma$ and $\bar{\gamma}_9$ becomes the same.
%\bea
%\Gamma_\pm \equiv \frac{1}{2} (1 \pm \Gamma) .
%\eea
%For $Spin(8)$ 
%the charge conjugation matrix 
%can be chosen as identity.
%If we choose such representation,
%the charge conjugation matrix for
%$Spin(1,10)$
%Clifford algebra
%can be chosen as
%\bea
%C = c \otimes 1
%\eea
%where $c$ is a charge conjugation matrix in $Spin(1,2)$.
%

\section{Majorana spinors}
\label{Majorana}
%Chirality condition in 8D direction:
%\bea
%\Gamma \Psi = - \Psi
%\eea
Majorana condition in 11D:
\bea
\Psi = C\bar{\Psi}^T .
\eea
Conjugate momentum (for kinetic terms the same to (\ref{BLaction}))
\bea
\Pi_{\Psi_\a}
=
\frac{i}{2}
(\bar{\Psi}\Gamma^0)_\a 
=
\frac{i}{2}
(\Gamma^{0}{}^T C^{-1}\Psi)_\a .
\eea
Dirac bracket:
\bea
\{\Psi_\a , \Psi_\b \}_D 
= 
- i 
\left(
\Gamma_-
\Gamma^{0} 
%\frac{1-\Gamma}{2} 
C
\right)_{\a\b} ,
\eea
where here and in the following we
suppress space coordinates and spinor indices
when it is obvious.

\section{Supercharge commutation relations}
\label{qq}

\bea
\tq = - \Gamma^0 \Psi_\C ,
\eea
\bea
\ql &=&  
-
\la
\Gamma^I\Psi, D_0 X^I
\ra
-
\la
\Gamma^I\Gamma^0\Gamma^i\Psi,
D_i X^I
\ra
\nn \\
&&
-
\frac{1}{6}
\la
\Gamma^{IJK}\Gamma^0 \Psi, [X^I,X^J,X^K]
\ra .
\eea
For any fields $\Phi$ in the BLG model, 
\bea
i \{ \bar{\eta}_\a \tQ_\a,\Phi \}_D
=
\delta_{\eta} \Phi,
\eea
\bea
i \{ \bar{\e}_\a Q_\a,\Phi \}_D
=
\delta_\e \Phi.
\eea
We obtain
\bea
i
\{\tq_\a, \Psi_{\C \b} \}_D
=
\left(
%\frac{1 - \Gamma}{2}
\Gamma_-
C
\right)_{\a\b}.
\eea
Dirac brackets for super charges:
\bea
i
\{\tq_\a, \tQ_\b  \}_D 
=
\left(
%\frac{1 - \Gamma}{2} 
\Gamma_-
\Gamma^0 C 
\right)_{\a\b}  ,
\eea
\bea
&&i \{\tq, \QL \}_D + i \{ \ql, \tQ \}_D \nn \\
&=&
 \Gamma^I C \pa_0 X^I_\C
- \ve^{0i}{}_{j} \Gamma^j \Gamma^I C \pa_i X^I_\C
\nn \\
&-&
\frac{1}{12}
\ve^{0}{}_{ij}
\la
[X^I,X^J,X^K],
T_\C
\ra
\Gamma^{ijIJK} C  .
\eea
This leads to (\ref{tQtQ}).
\bea
&& i \{\ql , \QL \}_D \nn \\
&=&
2 p_\mu
%\frac{1+\Gamma}{2}
\Gamma_+
\Gamma^\mu C 
\nn \\
&-&
\la
D_i X^I , D_j X^J
\ra
\ve^{0ij}
%\frac{1+\Gamma}{2} 
\Gamma_+
\Gamma^{IJ}
C
\nn \\
&+& 
\la
D_0 X^I , [X^I, X^J, X^K]
\ra
%\frac{1+\Gamma}{2} 
\Gamma_+
\Gamma^{JK}
C
\nn \\
&+&
\frac{1}{3}
\la
D_i X^I, [X^J, X^K, X^L]
\ra
%\frac{1+\Gamma}{2} 
\Gamma_+
\Gamma^{0iIJKL}  C
\nn \\
&-&
\frac{1}{4}
\la
 [X^I, X^J, X^K],[X^I, X^L, X^M]
\ra
%\frac{1+\Gamma}{2} 
\Gamma_+
\Gamma^{0JKLM}
C  .
\eea
This leads to (\ref{QLQL}).

The bosonic part of the energy-momentum tensor:
\bea
T_{\mu\nu}
=
\la
D_\mu X^I, D_\nu X^I
\ra
-
\eta_{\mu\nu}
\left(
\frac{1}{2}
\la
D^\rho X^I, D_\rho X^I  
\ra
+
V(X)
\right),
\eea
The bosonic part of the Hamiltonian density:
\begin{eqnarray}
{\cal H} &=& 
 \frac{1}{2}\la P^I, P^I \ra
+\sum_{i =1,2}
\frac{1}{2} \la D_i X^I , D_i X^I \ra
+ V (X).
\end{eqnarray}
%where
%\begin{equation}
%V (X) = \frac{1}{12}
%\langle [X^I, X^J, X^K],
%[X^I, X^J, X^K]
%\rangle.
%\end{equation}
The momentum densities:
\bea
p_0 =  {\cal H}, \quad
p_i = 
\la D_0 X^I , D_i X^I \ra, \quad
p_I = \pa_0 X^I_\C  .
\eea

%%%%%%%%%%%%
%References
%%%%%%%%%%%%%%%%%%%%%%%%
\bibliography{M2refs}
\bibliographystyle{utphys}
%%%%%%%%%%%%%%%%%%%%%%%%

\end{document}